# Atomic-scale Origin of Interfacial Dzyaloshinskii–Moriya interaction in Pt/Fe/Au

Emanuele Longo[1,*], Gianluca Gubbiotti[2], Matteo Belli[3], Claudia Wiemer[4], Mario Alia[4], M. Fanciulli[5], Roberto Mantovan[4,**]

1. Institut de Ciència de Materials de Barcelona (ICMAB-CSIC), Campus UAB, Bellaterra, Catalonia 08193, Spain
2. Cnr-Istituto Officina dei Materiali, Unità di Perugia, Via A. Pascoli 06123, Italy
3. CNR-IMEM Unit of Parma, Parco area delle Scienze 37/A, 43124 Parma, Italy
4. CNR-IMM, Unit of Agrate Brianza (MB), Via C. Olivetti 2, 20864, Agrate Brianza (MB), Italy
5. Department of Chemistry, University of Torino, Via P. Giuria 9, 10125, Torino, Italy

*elongo@icmab.es **roberto.mantovan@cnr.it

**Abstract**

Recent results show remarkable interfacial-Dzyaloshinskii-Moriya interaction (i-DMI) effect in magnetic heterostructures composed of relatively thick Fe thin films ($\gg 1\ nm$). To investigate its origin, we present a thorough magnetic characterization of the $SiO_2$/$^{57}$Fe(t)/Au and $SiO_2$/Pt/$^{57}$Fe(t)/Au heterostructures (t = 1.6 - 2.2 nm). An i-DMI constant $D_s^{Pt} = 0.43\ \mathrm{mJ/m^2}$ is extracted for the Pt/Fe/Au system, as probed by wavevector-resolved Brillouin light scattering spectroscopy. Ferromagnetic resonance and conversion electron Mossbauer spectroscopy provide evidence that significant surface anisotropy is present in Pt/Fe/Au and an interfacial Fe fraction (~18%) exhibits tilted magnetization (~25° out of plane) and enhanced hyperfine field, correlating with the stronger i-DMI.

**Introduction**

The research of novel functional materials and physical paradigms is nowadays fundamental to fulfill the increasing request of highly performing electronic devices (i.e., ultra-fast, low power consumption).[1,2] In the field of spintronics, it was largely demonstrated that the use of ferromagnetic materials (FM) (i.e., Fe, Co, CoFeB) offers the possibility to exploit a variety of physical phenomena for both memory storage and logic applications (e.g., MRAM, Datta-Das transistor, STT-MRAM).[3] In this context, the study of the interfaces between materials covers a central role to induce and control specific physical phenomena.[4] This is the case of the interfacial Dzyaloshinskii–Moriya interaction (i-DMI), a phenomenon attracting huge interest for the stabilization of non-conventional magnetic structures with helical magnetic order (e.g., magnetic bubbles, spin spirals, chiral domain walls, skyrmions) that can be exploited for the implementation of the future logic and memory storage applications.[5,6] The i-DMI originates from the breaking of the inversion symmetry at the interface between magnetic and high spin orbit coupling (SOC) materials, favoring non-collinear spin structures. A technologically relevant aspect is the search of heterostructures characterized by a large i-DMI effect, being a strict requirement to create smaller magnetic structures and for enhancing the velocity of their motion under the application of external stimuli (i.e., electric field). With this aim, FMs with strong perpendicular magnetic anisotropy (PMA) are usually required, characteristics present at sub-nanometric thickness regimes.[7] Nevertheless, the use of in-plane magnetized FM films of few nanometers was recently demonstrated to be a viable solution to obtain heterostructures with remarkable i-DMI,[8,9] despite the lack of a clear description of the micromagnetic origin of such an effect.

In this contribution, we investigate the chemical, structural, magnetic properties of two representative Fe-based heterostructures, namely $SiO_2$/Pt/Fe/Au and $SiO_2$/Fe/Au. In particular, our interest is in identifying the potential presence of i-DMI and to understand its origin at the most atomic-scale. First, a comprehensive characterization of the static and dynamic magnetic properties was carried out in both systems by combining broadband ferromagnetic resonance (BFMR) and vibrating sample magnetometry (VSM). Subsequently, Brillouin light scattering (BLS) spectroscopy, a powerful and straightforward technique for measuring the DMI constant through the nonreciprocal spin-wave propagation, was employed to determine the i-DMI constant in the investigated films.[10] While no i-DMI is detected in the $SiO_2$/Fe/Au sample, a significant i-DMI was observed in the Pt/Fe/Au system. Obtaining such a clear i-DMI response in a relatively thick Fe layer is quite unexpected. We propose that this effect originates from the presence of a fraction of Fe atoms in the Pt/Fe/Au heterostructure exhibiting an average out-of-plane magnetic moments' orientation of approximately 30°, as revealed by conversion electron Mössbauer spectroscopy (CEMS). Notably, this scenario is absent in the $SiO_2$/Fe/Au sample.

**Experimental**

The samples investigated in this study are Si//$SiO_2$(100 nm)/Pt(15 nm)/$^{57}$Fe(1.6 nm)/Au(4 nm) and Si//$SiO_2$(100 nm)/$^{57}$Fe(2.2 nm)/Au(4 nm), hereafter referred to as Pt/Fe/Au and $SiO_2$/Fe/Au, respectively. The samples thicknesses are extracted by X-ray reflectivity (XRR) measurements. XRR spectra were collected with a commercial scintillator and the data simulated with a matrix formalism model corrected by a Croce-Nevot factor (see Supplementary Information).[11] The Pt layer was deposited via DC magnetron sputtering, while the $^{57}$Fe/Au bilayer was sequentially e-beam evaporated onto both the reference $SiO_2$(100 nm) substrate and the $SiO_2$(100 nm)/Pt(15 nm) sample.

The use of isotopically enriched $^{57}$Fe enables CEMS, providing atomic-scale insight into the chemical, structural, and magnetic properties of the Fe atoms within both heterostructures. CEMS measurements were performed using a constant-acceleration drive, with the sample mounted as an electrode in a parallel-plate avalanche counter filled with acetone gas.[12] An α-$^{57}$Fe foil was used for the CEMS velocity scale calibration and all the reported isomer shifts are relative to α-Fe at room temperature. CEMS data analysis was carried out with the Vinda program.[13] A 25 mCi 57Co source in a Rhodium matrix was employed and operated in a constant acceleration mode.

Brillouin light scattering (BLS) spectroscopy of thermally excited spin waves was employed to quantify the strength of the i-DMI, taking advantage of the wave-vector resolution of this technique. The BLS measurements were performed by focusing approximately 200 mW of monochromatic light from a solid-state laser (λ = 532 nm) onto the sample surface. The backscattered light was analyzed using a Sandercock-type (3 + 3)-pass tandem Fabry–Perot interferometer.[14] During the experiments, a bias magnetic field of H = 200 mT was applied parallel to the sample surface, while the in-plane spin-wave wave vector $k$ was swept along the perpendicular direction (Damon–Eshbach configuration). Due to the momentum conservation law in the photon–magnon scattering process, the in-plane wave vector is directly related to the incidence angle of the probing light ($\theta_i$) according to

$$k = \frac{4\pi}{\lambda} \sin\theta_i \qquad [1]$$

In our measurements, $k$ was systematically varied in the range from 0 to $2.04 \times 10^7$ rad/m, providing access to the DE spin wave with different propagation vectors.

Broadband ferromagnetic resonance (BFMR) measurements were carried out using an Anritsu MG3694C source (1 – 40 GHz), connected to a grounded coplanar waveguide (GCPW). The samples were mounted in a flip-chip configuration, with the ferromagnetic film facing the GCPW and electrically insulated by a 75 µm Kapton spacer to avoid short-circuiting the signal line. The sample–GCPW assembly was placed between the pole pieces of a Bruker ER-200 electromagnet, with the film plane aligned parallel to the external magnetic field $H_{ext}$ (in-plane configuration). During measurements, an RF signal of fixed frequency was injected into the GCPW, and the transmitted signal was rectified using a diode to produce a DC voltage, which was then measured with a lock-in amplifier.

The magnetic hysteresis loops were recorded using vibrating sample magnetometry (VSM) in a Quantum Design Physical Property Measurement System (PPMS).

**Results and Discussion**

Figure 1a and 1b display the BLS measurements performed on the $SiO_2$/Fe/Au and the Pt/Fe/Au heterostructures, respectively, for a wave vector $k = 1.81 \times 10^7 rad/m$ (i.e., $\theta_i$ = 50 degrees). The empty red symbols indicate the Stokes shift in both panels, while the blue full symbols the anti-Stokes ones. The solid lines are the Lorentzian fit of the reported datasets, from which the frequency corresponding to the Stokes ($f_S$) and anti-Stokes ($f_{AS}$) signals are extracted. The $\Delta f = |f_S| - |f_{AS}|$ is a relevant quantity which is proportional to the strength of the i-DMI,[15] and obey to the following relation

$$\Delta f = \frac{2\gamma D_s}{\pi M_s} k \qquad [2]$$

with $\gamma$ representing the gyromagnetic ratio, $M_s$ the saturation magnetization and $D_s$ a parameter quantifying the strength of the i-DMI.

In $SiO_2$/Fe/Au we measure $\Delta f = 0.15 \pm 0.06$ GHz, slightly higher than the uncertainty in the determination of the BLS peaks (~ 0.1 GHz), while for Pt/Fe/Au $\Delta f = 0.62 \pm 0.06$ GHz. These values are in full agreement with the ones extracted in Ref.[8] for heterostructures with the same materials composition, thus confirming the reliability of the obtained results.

To quantify the strength of the i-DMI in the Pt/Fe/Au sample with high accuracy, we plotted the $\Delta f$ vs $k$ dependence, as shown in Figure 1c (black spheres), and applied the same procedure which has been used for the $SiO_2$/Fe/Au sample. The red solid line indicates the fit using Eq.2. In order to extract the value of the $D_s$ parameter related to the i-DMI intensity, $\gamma$ and $M_s$ should be known.

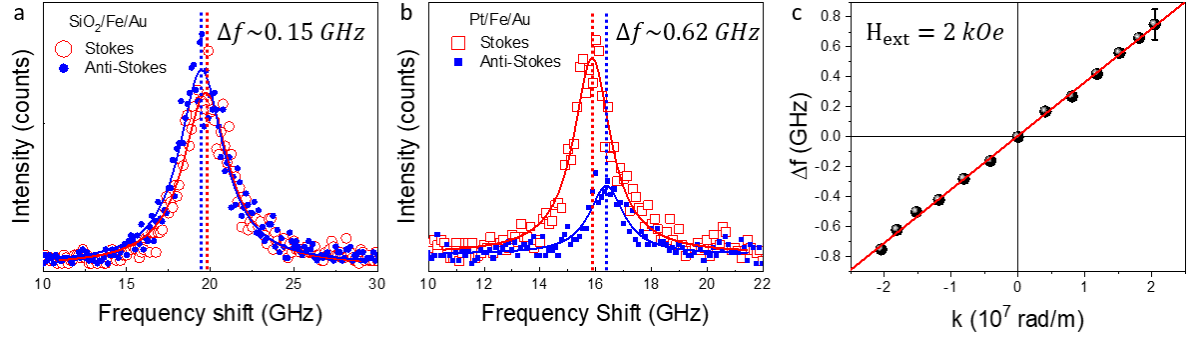

**Figure 1**: BLS spectra on the SiO$_2$/Fe/Au and Pt/Fe/Au heterostructures. (a) The empty red circles and the full blue circles represent the Stokes and anti-Stokes peaks for sample SiO$_2$/Fe/Au, with a small frequency difference of $\Delta f = 0.15\ GHz$ is measured between the two peaks at k=1.81×10$^7$ rad/m. Similarly, the Pt/Fe/Au sample data are shown, and the solid lines represent their Lorentzian fits. In this case, a frequency shift $\Delta f = 0.62\ GHz$ is observed between the Stokes and anti-Stokes BLS peaks. (c) Frequency versus k-vector curve measured under an external field H$_{ext}$ = 2 kOe for the Pt/Fe/Au sample. The red solid line is the fit of the dataset using Eq. 2.

To have access to the latter quantities we performed BFMR measurement on the SiO$_2$/Fe/Au and the Pt/Fe/Au samples, providing their magnetodynamic response. Figure 2a displays the evolution of the Kittel curves in the in-plane (IP) configuration for SiO$_2$/Fe/Au (black diamonds) and for Pt/Fe/Au (blue triangles) samples, where the RF-frequency exciting the system $f_{res}$ in plotted as a function of the resonant magnetic field $H_{res}$. The red solid lines represent the data fit using the Kittel model for the IP geometry,[16] which is described by the following Eq. 3.

$$f_{res} = \frac{\gamma}{2\pi}\sqrt{H_{res}\left(H_{res} + 4\pi M_{eff}\right)} \qquad (3)$$

where $M_{eff}$ the effective magnetization, $\gamma = g\frac{e}{2m_e}\left[\frac{Hz}{Oe}\right]$, where $e$ and m$_e$ are the charge and the mass of the electron, and $g$ is the Landè g-factor.[16] The experimental $f_{res}(H_{res})$ dataset is shown in Figure 2a and fitted using Eq. (3). To minimize the uncertainty on the extracted values of $\gamma$ and $M_{eff}$, we adopted an iterative fitting procedure. This approach consists of repeatedly fitting the data in Figure 2a while progressively increasing the upper limit of the frequency range (denoted as $f_{top}$), following the method described in Ref.[17]. By plotting the resulting $\gamma$ and $M_{eff}$ values as a function of $f_{top}$, an exponential trend is observed, which allows us to extract their asymptotic values. These trends are reported in Figures 2b and 2c, respectively. The gyromagnetic ratio $\gamma$ turned out to be the same for both samples, with a value $(1.88 \pm 0.02) \cdot 10^7 \frac{Hz}{Oe}$. Conversely, $M_{eff}$ differs for the two samples of almost 50%, being $M_{eff}^{SiO2} = 543\ \pm 46\ \frac{emu}{cm^3}$ and $M_{eff}^{Pt} = 1012\ \pm 112\ \frac{emu}{cm^3}$.

$M_{eff}$ accounts for magnetic anisotropy effects present in the magnetic heterostructure, and it can be written as $4\pi M_{eff} = 4\pi M_s - H_k \propto 4\pi M_s - \frac{2K_s}{M_s t_{Fe}}$, where $M_s$ is the saturation magnetization, $t_{Fe}$ the thickness of the Fe layer, $H_k$ and $K_s$ the magnetic anisotropy field and the surface magnetic anisotropy constant, respectively.

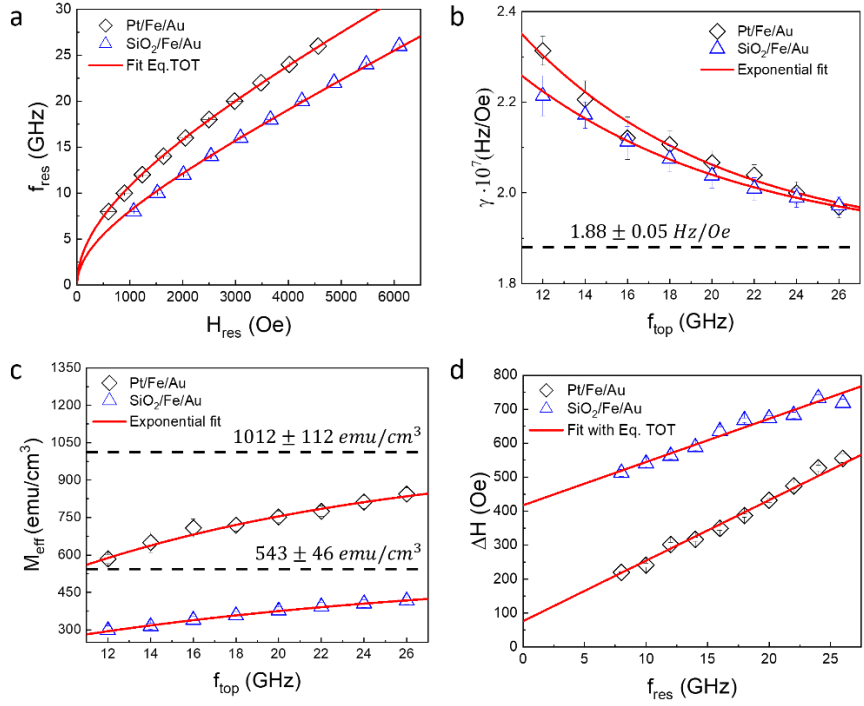

**Figure 2**: (a) Evolution of the Kittel curves for sample SiO$_2$/Fe/Au (blue empty triangles) and the Pt/Fe/Au (black empty diamonds). Panels (b) and (c) represent the asymptotic values for $\gamma$ and $M_{eff}$ as extracted by the iterative procedure described in the main text. (d) Dispersion of the FMR signal linewidth for the same samples of panel (a), from which the damping constant value $\alpha$ is extracted.

In Figure 2d, the linewidth $\Delta H$ of the BFMR signal is plotted as a function of $f_{res}$ and, exploiting Eq. 5, the values of the damping constant $\alpha$ and the inhomogeneous broadening $\Delta H_0$ can be found.

$$\Delta H = \Delta H_0 + \frac{4\pi}{|\gamma|} \alpha f_{res} \qquad (5)$$

From the linear fit (red solid lines) of the datasets reported in Figure 2d, we obtained $\alpha^{SiO2} = (19 \pm 1) \cdot 10^{-3}$, $\Delta H_0^{SiO2} = 417 \pm 17\ Oe$, $\alpha^{Pt} = (27 \pm 1) \cdot 10^{-3}$ and $\Delta H_0^{Pt} = 76 \pm 6\ Oe$. A clear enhancement of $\alpha$ is originated by the Pt layer, which is attributed to the spin pumping effect from the Fe layer into the Pt one, acting as a spin sink.[18–20] $\Delta H_0$ acts as an indicator of the magneto-structural quality of the FM Fe layer, accounting for magnetic imperfections (e.g., magnetic dead layers) or texturing. For that, the considerably higher $\Delta H_0^{SiO2}$ with respect to $\Delta H_0^{Pt}$ is a fingerprint of the lower magnetic quality of the Fe layer when grown on SiO$_2$ substrates.

To evaluate the role of PMA in promoting i-DMI and in the studied systems, the knowledge of $K_s$ is crucial. Thus, we performed VSM measurement to extract the $M_s$ value for both samples, a parameter also necessary to finally quantify the strength of the i-DMI (see Eq. 1). Figure S2a and S2b display the hysteresis loop acquired with the external magnetic field (H$_{ext}$) positioned IP (red squares) and OOP (green triangles) (see Supplementary Information). The shape of the hysteresis loop shows that the Fe layer has an IP easy plane for both samples, with saturation values $M_s^{SiO2} = 820\ \frac{emu}{cm^3}$ and $M_s^{Pt} = 1500\ \frac{emu}{cm^3}$. If compared with the $M_s$ for bulk Fe ($\sim 1710\ \frac{emu}{cm^3}$),[21] the latter values indicate that the Fe layer grown on Pt has higher magnetic order with respect to the one grown on SiO₂, fully in accordance with the BFMR results (see Figure 2d). By using the measured $M_s$ values, the shape anisotropy constant $K_{shape} = -\frac{1}{2}\mu_0 M_s^2$ can be calculated for the two heterostructures, being $K_{shape}^{SiO2} = 4.2 \cdot 10^6 \frac{erg}{cm^3}$ and $K_{shape}^{Pt} = 14 \cdot 10^6 \frac{erg}{cm^3}$. We also quantified the $K_s$ constants as discussed above, $K_s^{SiO2} = 0.31 \frac{erg}{cm^2}$ and $K_s^{Pt} = 0.74 \frac{erg}{cm^2}$, in agreement with previously reported values for Fe-metal interfaces.[22,23] The sign of $K_s$ indicates that both samples possess an energy contribution favoring the out-of-plane (OOP) orientation of the Fe magnetization vector, even though the overall sample magnetization lies within the film plane due to the dominant magnetostatic energy. The OOP anisotropy is more pronounced in the sample with the Pt layer.

Finally, substituting the correspondent parameters in Eq.1, the i-DMI strength can be calculated for both samples, obtaining $D_s^{SiO2} = 0.057$ mJ/m² and $D_s^{Pt} = 0.43$ mJ/m². The more than eight times lower i-DMI observed for the SiO₂/Fe/Au system can be interpreted by the fact that SiO₂ is a light-element with negligible SOC, and it is also an electrical insulator. While the interface still breaks inversion symmetry, there is no heavy element present to provide the strong spin-SOC channel needed to generate a robust i-DMI. The SiO₂/Fe interface thus lacks the orbital hybridization and spin–orbit scattering that are essential to establish a significant chiral exchange. As a result, the i-DMI in SiO₂/Fe is extremely weak, effectively negligible compared to Pt/Fe. In order to investigate on the chemical-structural properties and the magnetic nature of the studied samples, CEMS measurements are performed. CEMS spectra in Fig. 3 (a) and (b) appear clearly different. Looking to the raw data (blue points), both are characterized by well-defined magnetically-split sextets. The hyperfine parameters extracted from the fit of the CEMS data in both samples are summarized in Table 1. The relative area fractions in Table 1 are obtained by assuming the same Debye-Waller factor $f$=1 for all the components. For interpreting the data in Table 1, we start from some common features appearing in both spectra, namely the DOUB and DIST components. As clearly seen in Table 1, the hyperfine parameters are very close for these components in both $SiO_2$/Fe/Au and Pt/Fe/Au samples, suggesting they may have a common origin. Due to close vicinity with the DOUB's hyperfine parameters with those obtained to interpret a paramagnetic doublet in Si(111)/$^{57}$Fe(5nm)/Au(5nm) sample,[24] we attribute this component to Fe atoms at the interface with the top Au layers, in both the $SiO_2$/Fe/Au and Pt/Fe/Au samples. Apart from paramagnetic Fe-Au intermetallics, several magnetic phases have been observed by Mössbauer spectroscopy at mixed Fe-Au systems, with a large variety of hyperfine parameters depending on the relative Fe-Au concentration.[25] Certainly, at least a fraction of this DIST component may be ascribed to Fe-Au magnetic interaction. Additionally, by considering the higher relative contribution of DIST in $SiO_2$/Fe/Au than in Pt/Fe/Au (see Table 1), we think that some fraction of Fe atoms coordinating in the DIST configuration on the $SiO_2$ substrate, transform into a more ordered magnetic configuration in Pt/Fe/Au, as due to the new Fe-2 component appearing there (Fig. 3 and Table 1). At least in $SiO_2$/Fe/Au, a fraction of DIST could be attributed to "inner" Fe atoms within the 2 nm thickness (i.e., not necessarily at the interface with $SiO_2$ and/or Au), as due to the growth of a relatively poor-quality Fe layer on top of $SiO_2$, in accordance with BFMR and VSM results (see Figures 2d and 3a). On the other hand, the largest part of Fe atoms in $SiO_2$/Fe/Au are Fe atoms coordinated in the Fe-1 fraction, which accounts for those organized in a more ordered magnetic configuration, even if $B_{hf}$ is still lower than that in cubic bulk α-Fe. The nature of Fe-1 appears to be very similar in both $SiO_2$/Fe/Au and Pt/Fe/Au as supported by the very similar hyperfine parameters obtained from the fit of the data (Table 1).

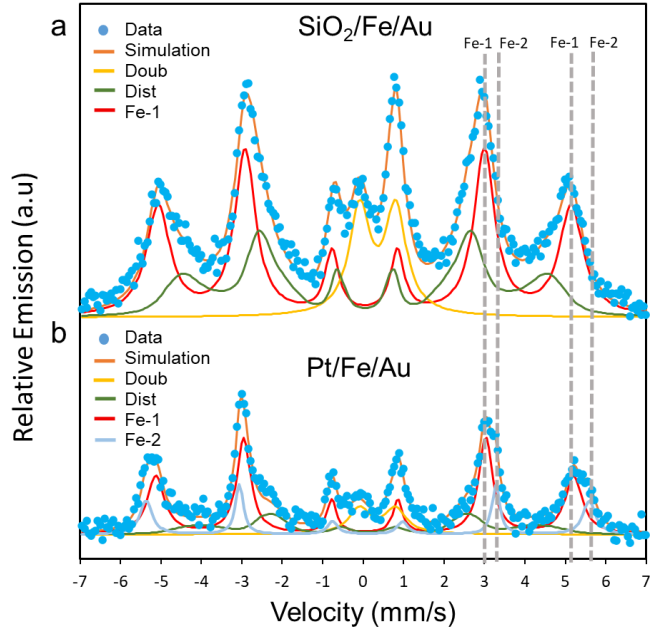

**Figure 3**: CEMS of (a) SiO₂/Fe/Au and (b) Pt/Fe/Au samples. In both panel, blue dots are the experimental data while solid lines constitute the fitting components (see insets) and the total spectra simulation (solid orange line). The gray dotted lines are "guide for the eyes" to appreciate the clear need for an additional magnetically-split sextet (Fe-2) to interpret the data in (b), while the common Fe-1 component is alone providing a good interpretation for the magnetically-split signal in (a).

**Table 1.** Hyperfine parameters for all the spectral components used to interpret the CEMS data shown in Figure 3, for both SiO2/Fe/Au and Pt/Fe/Au samples.

| | Hyperfine parameters | Fe-1 | Fe-2 | DOUB | DIST |
|---|---|---|---|---|---|
| **SiO$_2$/Fe/Au** | $<B_{hf}>$ (T) | 31.63(2) | . | - | 26.2(5) |
| | $\delta$ (mm/s) | 0.039(2) | - | 0.360(4) | 0.049(5) |
| | $\Delta E_Q$ (mm/s) | 0 | - | 0.891(7) | 0 |
| | $<\Gamma>$ (mm/s) | 0.58(3) | - | 0.636(9) | 0.48(5) |
| | x | 4 | - | - | 4 |
| | Area (%) | 52(3) | - | 16(3) | 32(2) |
| **Pt/Fe/Au** | $<B_{hf}>$ (T) | 32.00(2) | 33.93(3) | - | 25.7(5) |
| | $\delta$ (mm/s) | 0.040(2) | 0.110(4) | 0.36(1) | 0.14(2) |
| | $\Delta E_Q$ (mm/s) | 0 | 0 | 0.89(2) | 0 |
| | $<\Gamma>$ (mm/s) | 0.41(3) | 0.31(6) | 0.64(3) | 0.5(1) |
| | x | 4 | 2.8(1) | - | 4 |

| | | | | |
|---|---|---|---|---|
| Area (%) | 53(2) | 18(2) | 11(2) | 19(2) |

The additional Fe-2 is uniquely able to interpret several aspects emerging in the CEMS data in Pt/Fe/Au when compared to SiO$_2$/Fe/Au, which are: (i) the clear asymmetry between lines 2 and 5, (ii) the enhanced hyperfine magnetic field that is represented by high-velocity side broadening of lines 1 and 6, and (iii) the emergence of the right-side shoulder in line 5. All these features can be satisfactorily justified by including this additional Fe-2 component in the CEM-spectrum of Pt/Fe/Au.

Very interestingly, Fe-2 differs from all other magnetically split components detected in both SiO$_2$/Fe/Au and Pt/Fe/Au, for what concerns the relative line ratio. The intensity ratio of the magnetically-split lines in a CEMS spectrum can be expressed as 3:x:1:1:x:3, with $\cos^2\Theta = (4-x)/(4+x)$, where $\Theta$ is the polar angle between the average direction of the Fe magnetic moments of that specific spectral component and the c-axis (i.e., normal to the film plane). This means that x=4 corresponds to magnetic moments fully lying in the film plane, while any other values indicate a Fe magnetic moment pointing along the OOP direction, with x=0 representing a full PMA. From Table 1, we see that Fe-2 (18% fraction, out of the total Fe atoms) in Pt/Fe/Au is characterized by an OOP fraction, with x=2.8 corresponding to $\Theta$= 65°, i.e., as 25° tilt from the in-plane direction.

As previously reported, in the case of strong PMA in Fe/Pt systems, the squared hysteresis loops for a magnetic field applied perpendicular to the film plane corresponds to the ideal 3:0:1:1:0:3 ratio in CEMS analysis.[26] On the other hand, this may not be the case in our system, where the relatively low OOP contribution we detect by CEMS in Pt/Fe/Au, indeed, is not reflected in "volume-sensitive" magnetic measurements such as BFMR and VSM, which detects an almost ideal IP magnetization (Fig.3) in both films.

To identify the possible location of the Fe atoms belonging to the Fe-2 phase it must be considered the correspondent $B_{hf}$ value, which turns out to be higher than Fe in α-Fe phase (Table 1). Indeed, among other systems, interfacial enhancement of $B_{hf}$ has been previously reported in Fe/Pd[27] and Fe/Pt.[28,29] This led us to propose that this 18.4% fraction of Fe atoms that are characterized by partly OOP magnetization (by 25°) are most likely located close to the interface with Pt, where they experience a slight OOP anisotropy. Considering an average value for a Fe monolayer (ML) of ∼2 Å (i.e., average of the interplanar distance of the (100), (110) and (111) planes in a Fe body-centred cubic structure), in the Pt/Fe/Au sample there are 8 MLs. Thus, the detected 18% fraction of the Fe-2 phase atoms corresponds to about 1.5 ML, likely located at the interface with Pt. Based on the adopted analysis, we suggest that these Fe atoms experience a strong interface OOP anisotropy at zero magnetic field, which can be at the origin of the much higher i-DMI constant extracted for Pt/Fe/Au when compared to SiO$_2$/Fe/Au.

The simultaneous existence of Fe-1 (and DIST) and Fe-2, as proposed based on CEMS measurements, may induce pronounced non-collinear magnetization effects (close to the Fe/Pt interface), potentially acting in synergy with SOC, thus activating the observed relatively large i-DMI in the Pt/Fe/Au sample.

**Conclusion**

In this work, we have thoroughly characterized the magnetic and atomic-structural properties of Fe-based heterostructures grown on Pt and on SiO$_2$, showing that a measurable interfacial i-DMI is present even in Fe films of relatively large thickness (~1.6–2.2 nm). Using BLS we detected a significantly larger i-DMI constant in Pt/Fe/Au compared to the SiO$_2$/Fe/Au reference, with extracted DMI constant of $D_S^{Pt} = 0.43$ mJ/m$^2$. and $D_S^{SiO2} = 0.057$ mJ/m$^2$, respectively. While this result could be expected due to the heavy underlaying Pt, it is in general not straightforward to be observed at the relatively high thicknesses above 1-2 nm due to the interfacial nature of i-DMI. The results confirm that a heavy-metal underlayer such as Pt significantly enhances the i-DMI when in contact with Fe. BFMR performed on the two systems revealed a higher crystalline quality when Fe is grown on Pt, leading to a significant increase of the surface anisotropy constant with respect to SiO$_2$ (i.e., $\frac{K_S^{Pt}}{K_S^{SiO2}} = 3.5$). CEMS analysis rationalized the microscopic origin of this difference: in Pt/Fe/Au we observe an ~18% fraction of Fe atoms characterized by an average magnetization tilted ~25° away from the film plane at remanence, together with an enhanced hyperfine field compared or higher than bulk-like Fe. We propose that this interfacial fraction (Fe-2), localized at the Pt/Fe interface, generates local anisotropy and electronic hybridization favorable to non-collinear magnetic moments, thus reinforcing the observed i-DMI. Overall, our results indicate that, beyond the simple surface-to-volume scaling $D_{eff} \propto D_s/t$, the atomic-scale interfacial microstructure (i.e., fraction of tilted Fe spins, modulation of $B_{hf}$ field, film crystalline quality) plays a crucial role in determining the strength of i-DMI in nanometer-thick Fe films.

**Funding**

E. L. acknowledges financial support of the Spanish Ministry of Science and Innovation through Projects PID2023-152225NB-I00 and Severo Ochoa MATRANS42 (CEX2023-001263-S), supported by MICIU/AEI/10.13039/501100011033 and FEDER, EU. E.L. also acknowledge support from Projects TED2021-129857B-I00 and PDC2023-145824-I00, funded by MCIN/AEI/10.13039/501100011033 and the European Union NextGeneration EU/PRTR, as well as from project 2021 SGR 00445 funded by the Generalitat de Catalunya. R. M. acknowledges financial support within the Horizon 2020 project SKYTOP "Skyrmion-Topological Insulator and Weyl Semimetal Technology" (FETPROACT-2018-01, n. 824123) and the PNRR MUR project PE0000023-NQSTI. G. G. acknowledge the European Union— Next Generation EU under the Italian Ministry of University and Research (MUR) National Innovation Ecosystem grant ECS00000041—VITALITY. CUP: B43C22000470005.

**Author Contributions**

E.L. performed the BFMR measurements and analyzed the data. E.L and M.B. developed the BFMR measurement setup. M.A. grew the samples. G.G. performed the BLS measurement and analyzed the data. M.F. supervised the BFMR measurements. R.M. performed and analyzed the CEMS data and coordinated the overall research activity. E. L. wrote the manuscript with contributions from all the co-authors.

**Conflict of interest**

The authors declare no conflict of interest.

**Data availability**

Data will be made available on reasonable request.

**Bibliography**

[1]     E. Commission, *https://ec.europa.eu/commission/presscorner/detail/en/QANDA_22_6229* **2022**.
[2]     N. Jones, *Nature* **2018**, *561*, 163.
[3]     X. Liu, K. H. Lam, K. Zhu, C. Zheng, X. Li, Y. Du, C. Liu, P. W. T. Pong, *IEEE Trans Magn* **2019**, *55*, DOI 10.1109/TMAG.2019.2927457.
[4]     R. Mantovan, T. Kampfrath, C. Ciccarelli, *Adv Mater Interfaces* **2022**, *9*, 2008.
[5]     K. K. Mishra, A. H. Lone, S. Srinivasan, H. Fariborzi, G. Setti, *Appl Phys Rev* **2025**, *12*, DOI 10.1063/5.0223004.
[6]     H. Vakili, W. Zhou, C. T. Ma, S. J. Poon, M. G. Morshed, M. N. Sakib, S. Ganguly, M. Stan, T. Q. Hartnett, P. Balachandran, J. W. Xu, Y. Quessab, A. D. Kent, K. Litzius, G. S. D. Beach, A. W. Ghosh, *J Appl Phys* **2021**, *130*, DOI 10.1063/5.0046950.
[7]     M. Kuepferling, A. Casiraghi, G. Soares, G. Durin, F. Garcia-Sanchez, L. Chen, C. H. Back, C. H. Marrows, S. Tacchi, G. Carlotti, *Rev Mod Phys* **2023**, *95*, DOI 10.1103/RevModPhys.95.015003.
[8]     W. Zhang, B. Jiang, L. Wang, Y. Fan, Y. Zhang, S. Y. Yu, G. B. Han, G. L. Liu, C. Feng, G. H. Yu, S. S. Yan, S. Kang, *Phys Rev Appl* **2019**, *12*, DOI 10.1103/PhysRevApplied.12.064031.
[9]     W. J. Zhang, F. Wei, B. Liu, Y. Zhou, S. S. Kang, B. Sun, *Chinese Physics B* **2024**, *33*, DOI 10.1088/1674-1056/ad1b41.
[10]    S. Tacchi, R. E. Troncoso, M. Ahlberg, G. Gubbiotti, M. Madami, J. Åkerman, P. Landeros, *Phys Rev Lett* **2017**, *118*, DOI 10.1103/PhysRevLett.118.147201.
[11]    E. Longo, C. Wiemer, R. Cecchini, M. Longo, A. Lamperti, M. Fanciulli, R. Mantovan, E. Longo, M. Fanciulli, A. Khanas, A. Zenkevich, *J Magn Magn Mater* **2019**, *474*, 632.
[12]    P. Gütlich, E. Bill, A. X. Trautwein, *Mössbauer Spectroscopy and Transition Metal Chemistry: Fundamentals and Applications* **2011**, 1.
[13]    H. P. Gunnlaugsson, *Hyperfine Interact* **2016**, *237*, 13.
[14]    G. Carlotti, G. Gubbiotti, *J. Phys.: Condens. Matter* **2002**, *14*, 8199.
[15]    A. K. Chaurasiya, C. Banerjee, S. Pan, S. Sahoo, S. Choudhury, J. Sinha, A. Barman, *Sci Rep* **2016**, *6*, DOI 10.1038/srep32592.
[16]    M. Farle, *Reports on Progress in Physics* **1998**, *61*, 755.
[17]    J. M. Shaw, H. T. Nembach, T. J. Silva, C. T. Boone, *J Appl Phys* **2013**, *114*, DOI 10.1063/1.4852415.
[18]    S. Martín-Rio, A. Pomar, L. Balcells, B. Bozzo, C. Frontera, B. Martínez, *J Magn Magn Mater* **2020**, *500*, DOI 10.1016/j.jmmm.2019.166319.
[19]    E. Longo, C. Wiemer, M. Belli, R. Cecchini, M. Longo, M. Cantoni, C. Rinaldi, M. D. Overbeek, C. H. Winter, G. Gubbiotti, G. Tallarida, M. Fanciulli, R. Mantovan, *J Magn Magn Mater* **2020**, *509*, 166885.
[20]    S. Martin-Rio, C. Frontera, A. Pomar, L. Balcells, B. Martinez, *Sci Rep* **2022**, *12*, DOI 10.1038/s41598-021-04319-z.
[21]    L. Sun, Y. Hao, C. L. Chien, P. C. Searson, P. C. Searson, *IBM J Res Dev* **2005**, *49*, 79.
[22]    K. B. Urquhart, B. Heinrich, J. F. Cochran, A. S. Arrott, K. Myrtle, *J Appl Phys* **1988**, *64*, 5334.
[23]    B. Heinrich, Z. Celinski, J. F. Cochran, A. S. Arrott, K. Myrtle, *J Appl Phys* **1991**, *70*, 5769.
[24]    E. Longo, L. Locatelli, M. Belli, M. Alia, A. Kumar, M. Longo, M. Fanciulli, R. Mantovan, *Adv Mater Interfaces* **2021**, *2101244*, 2101244.
[25]    A. Błachowski, K. Ruebenbauer, J. Przewoźnik, J. Zukrowski, *J Alloys Compd* **2008**, *458*, 96.
[26]    T. Sato, T. Goto, H. Ogata, K. Yamaguchi, H. Yoshida, *J Magn Magn Mater* **2004**, *272–276*, E951.
[27]    B. R. Cuenya, W. Keune, D. Li, S. D. Bader, *Phys Rev B Condens Matter Mater Phys* **2005**, *71*, 064409.
[28]    R. Wu, L. Chen, N. Kioussis, *J Appl Phys* **1996**, *79*, 4783.
[29]     H. Lassri, M. Abid, R. Krishnan, A. Fnidiki, J. Teillet, *J Magn Magn Mater* **1997**, *172*, 61.

# Atomic-scale Origin of Interfacial Dzyaloshinskii–Moriya interaction in Pt/Fe/Au

Emanuele Longo[1,*], Gianluca Gubbiotti[2], Matteo Belli[3], Claudia Wiemer[4], Mario Alia[4], M. Fanciulli, Roberto Mantovan[4,**]

1. Institut de Ciència de Materials de Barcelona (ICMAB-CSIC), Campus UAB, Bellaterra, Catalonia 08193, Spain
2. Cnr-Istituto Officina dei Materiali, Unità di Perugia, Via A. Pascoli 06123, Italy
3. CNR-IMEM Unit of Parma, Parco area delle Scienze 37/A, 43124 Parma, Italy
4. CNR-IMM, Unit of Agrate Brianza (MB), Via C. Olivetti 2, 20864, Agrate Brianza (MB), Italy

*elongo@icmab.es **roberto.mantovan@cnr.it

*Keywords: Spintronics, Dzyaloshinskii–Moriya interaction, Ferromagnetic Resonance, Mössbauer spectroscopy*

## Supplementary Information

1. **X-ray reflectivity (XRR) measurements**

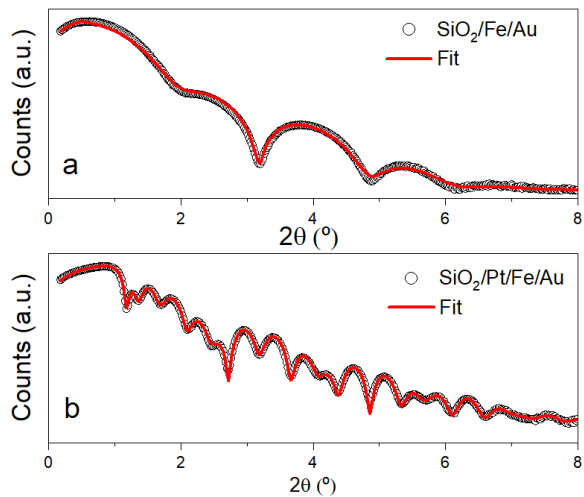

**Figure S1**: XRR curves acquired for the SiO$_2$/Fe/Au and SiO$_2$/Pt/Fe/Au samples (black circles). The red solid lines indicate the fit of the XRR data.

**Table S1**: XRR parameters as extracted from the fit reported in Figure S1. $\rho_e(e^-/Å^3)$ is the electronic density.

| Stack Figure S1a | Thickness (nm) | $\rho_e(e^-/Å^3)$ | Roughness (nm) |
|---|---|---|---|
| Au | 3,4 | 4.54 | 0,62 |
| Fe | 2,2 | 2.80 | 0,61 |
| SiO$_2$ | 50 | 0.73 | 0,35 |

| Stack Figure S1b | Thickness (nm) | $\rho_e(e^-/Å^3)$ | Roughness (nm) |
|---|---|---|---|
| Au | 4,1 | 4.71 | 0,52 |
| Fe | 1,6 | 2.10 | 0,58 |
| Pt | 14,8 | 5.18 | 0,49 |
| SiO$_2$ | 50 | 0.73 | 0,35 |

2. **Vibrating sample magnetometry (VSM) characterization**

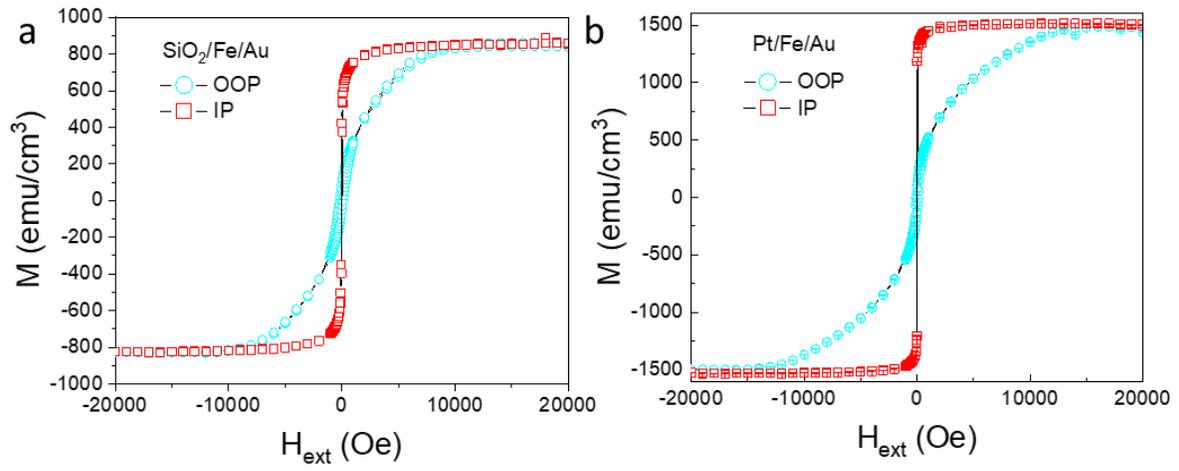

**Figure S2**: VSM measurements on $SiO_2$/Fe/Au and Pt/Fe/Au samples. In panel (a) and (b) are reported the hysteresis loops for the out-of-plane (OOP) (light blue circles) and in-plane (IP) (red squares) directions within the film plane.